\documentclass[amssymb,prb,twocolumn,showpacs]{revtex4}
\usepackage{graphicx}
\usepackage{epsfig}
\newfont{\bl}{cmbxsl10 scaled\magstep1}
\begin{document}
\title{Interlayer Exchange Coupling in (Ga,Mn)As-based Superlattices}
\author{P. Sankowski and P. Kacman} \affiliation{Institute of Physics,
  Polish Academy of Sciences, al.~Lotnik\'ow 32/46, 02-668 Warszawa, Poland}
\date{\today}
\begin{abstract}
  The interlayer coupling between (Ga,Mn)As ferromagnetic layers in
  all-semiconductor superlattices is studied theoretically within a
  tight-binding model, which takes into account the crystal, band and
  magnetic structure of the constituent superlattice components. It is
  shown that the mechanism originally introduced to describe the spin
  correlations in antiferromagnetic EuTe/PbTe superlattices, explains
  the experimental results observed in ferromagnetic semiconductor
  structures, i.e., both the antiferromagnetic coupling between
  ferromagnetic layers in IV-VI (EuS/PbS and EuS/YbSe) superlattices
  as well as the ferromagnetic interlayer coupling in III-V
  ((Ga,Mn)As/GaAs) multilayer structures. The model allows also to
  predict (Ga,Mn)As-based structures, in which an antiferromagnetic
  interlayer coupling could be expected.
\end{abstract}
\maketitle

Interlayer exchange coupling (IEC) -- the phenomenon, which was shown
to be responsible for the giant magnetoresistance effect,
\cite{parkin} and which led already to many applications of magnetic
metallic thin film structures \cite{prinz} -- was discovered in late
1980-s.  Since the first report on correlated magnetization vectors in
Fe/Cr/Fe trilayers, \cite{gruen} IEC was observed in a variety of
structures composed of metallic ferromagnetic (FM) layers separated by
nonmagnetic, metallic or insulating, spacer layers. The attempts to
explain this phenomenon were summarized in Ref.~\onlinecite{bruno}
where it was shown that IEC can be ascribed to the spin dependent
changes of the density of states resulting from the quantum
interference of conduction electron waves.

Although the FM and the metallic character of magnetic layers were
considered as inherent elements of the IEC effect, in 1995 the
interlayer spin correlations between antiferromagnetic (AFM) layers in
all-semiconductor superlattices (SL) were reported. \cite{gieb1} Next,
such coupling was also identified in semiconductor multilayer
structures with FM, (Ga,Mn)As \cite{akiba} and EuS \cite{kepa}, layers.
In addition to their basic science significance, these discoveries
were important because the all-semiconductor structures offer the
possibility to overcome the limitations brought about by the
technological incompatibility of FM metals and semiconductors.
Moreover, their properties can be easily controlled by temperature,
light or external electric fields.  From this applicational point of
view, the most interesting was the discovery of AFM coupling between
FM layers in EuS/PbS SL. \cite{kepa} In these structures, however, the
effect takes place only at very low temperatures -- bulk EuS is a
classical Heisenberg ferromagnet with the Curie temperature 16.6 K.
\cite{stachow} In (Ga,Mn)As-based FM structures, where a higher critical
temperature can be achieved, unfortunately only FM IEC was observed.
\cite{akiba,chiba,kepar,szusz,lee}

To explain the spin correlations observed in the AFM EuTe/PbTe and FM
EuS/PbS SL, a model was proposed, \cite{blin} in which the significant
role of the valence band electrons in IEC in all-semiconductor
magnetic/nonmagnetic layer structures was put in evidence. In
Ref.~\onlinecite{blin} it has been proven that quantum interference
between the spin-dependent perturbations in successive barriers, as
proposed by Bruno,\cite{bruno} is an effective mechanism for magnetic
long range correlations also when there are no free carriers in the
system. The IEC mediated by valence-band electrons, calculated within
this model, correlates antiferromagnetically the spins at the two
interfaces bordering each nonmagnetic layer of the SL.  Such spin-spin
interactions lead, in agreement with the experimental findings, to
zero net magnetic moment in the case of AFM EuTe/PbTe SL
\cite{kepaeute} and to an AFM coupling between successive FM EuS
layers in EuS/PbS SL.\cite{kepa,smits}  The strength of the obtained
IEC decreases rapidly (exponentially) with the distance between the
spins, i.e., with the thickness of the nonmagnetic spacer layer and
practically does not depend on the thickness of magnetic layers. In
Refs~\onlinecite{smits} and \onlinecite{story} a careful analysis of
the experimental results, in particular of the temperature and
magnetic field dependence of the SQUID magnetization, led the authors
to the conclusion that such IEC describes properly all the
neutron-scattering and magnetic observations in EuS/PbS structures
with ultrathin (ca $1.2$~nm thick) PbS spacers.  The traces of the
coupling observed by neutron scattering in samples with relatively
thick spacers were ascribed, however, to the weak but slowly decaying
contribution from the dipolar interactions. \cite{kepa,story,icps}

In the (Ga,Mn)As-based semiconductor ferromagnetic/nonmagnetic systems
interlayer coupling of opposite FM sign was observed -- by magnetic
measurements \cite{akiba,chiba,lee} and by neutron diffraction
\cite{szusz} and polarized neutron reflectometry.\cite{kepar} These
structures differ from the previously considered EuS/PbS multilayers
by many aspects, which all can affect the IEC.  First of all, in
contrast to the simple rock-salt crystal structure of EuS-based SL,
they crystallize in zinc blende structure. Moreover, PbS is a narrow
gap, whereas EuS is a wide gap semiconductor. In EuS/PbS SL the spacer
layers form deep wells in the energy structure of the multilayer -
here, the band structures of the magnetic ((Ga,Mn)As) and nonmagnetic
(GaAs, (Al,Ga)As) materials are either very similar or the spacer
layers introduce potential barriers for the carriers. It should be
noted, however, that in EuS-based structures the wider energy gap of
the spacer material does not lead to different character of IEC, but
results only in a reduction of the coupling' strength and range. This
was shown by theoretical studies of the coupling between EuS layers
separated by YbSe and SrS insulators \cite{sank} and confirmed by
neutron reflectivity experiments in EuS/YbSe SL.\cite{icm} Finally,
(Ga,Mn)As is not a magnetic but diluted magnetic semiconductor -- in
this ternary alloy a small, randomly distributed fraction of the Ga
cations is substituted by magnetic Mn ions. The spin splittings are
smaller than in EuS, the ferromagnetism is carrier-induced
\cite{dietl} and requires a considerable amount of free holes in the
valence band of the FM (Ga,Mn)As.

In Refs~\onlinecite{akiba} and \onlinecite{chiba} the observed much
weaker IEC in samples with high ($30\%$) Al content in the (Al,Ga)As
spacer led the authors to the conclusion that the coupling between the
FM layers is mediated by the carriers in the nonmagnetic layer.
Recently, it was also shown that introducing extra holes by Be-doping
of the GaAs spacer increases the interlayer coupling.\cite{lee} To
explain the spin correlations between (Ga,Mn)As layers the RKKY
mechanism and the models tailored for metallic systems were invoked.
\cite{jungwirth,boselli} In this paper, in order to describe the
spin-dependent band structure effects which can lead to IEC in
(Ga,Mn)As-based semiconductor SL, we built a tight-binding model in
the spirit of the approach used before for IV-VI semiconductor
magnetic multilayers.\cite{blin} The model was applied to the SL
consisting of alternating $m$ monolayers of (Ga,Mn)As, with the Mn
content 4{\%} or 6{\%}, and $n$ monolayers of GaAs, (Al,Ga)As or
GaAs:Be, i.e., to the structures studied experimentally.

In order to construct the empirical tight-binding Hamiltonian matrix
for the SL one has to describe first the constituent materials, to
select the set of atomic orbitals for every type of involved ions and
to specify the range of the ion-ion interactions.  In the following, we
assume that the proper description of SL band structure is reached
when the Hamiltonian reproduces in the $n=0$ and $m=0$ limits the band
structures of the constituent magnetic and nonmagnetic materials,
respectively.  Bulk GaAs is tetrahedrally coordinated cubic material
in which each cation (anion) is surrounded by four anion (cation)
nearest neighbors (NN) along the [1, 1, 1], [1,-1,-1], [-1, 1,-1] and
[-1,-1, 1] directions, at the distances $a\sqrt{3}/4$ (where $a=5.653
$~\AA \ is the lattice constant). GaAs is a nonmagnetic, direct gap
semiconductor with the valence band maximum at the center of the
Brillouin zone. The top of the valence band is formed by two twofold
degenerate p-bands. The third p-band is separated from the two by
spin-orbit splitting, $\Delta_{so}=0.34$ eV.  The band structure of
GaAs was described by many authors. Here we use the structure obtained
by Jancu {\it at al.}\cite{jancu} within $sp^3d^5s^*$ empirical
tight-binding model, which takes into account the $s$, $p$ and $d$
orbitals for both, anions and cations. As shown in
Ref.~\onlinecite{jancu}, the inclusion of $d$-orbitals improved
considerably the description of the band structure in the vicinity of
X-high symmetry point of the Brillouin zone. The spin-orbit
interactions were added to the model by including the contribution
from the $p$ valence states. The tight-binding model parameters were
obtained by fitting the on-site energies and the two-center NN
integrals in the Hamiltonian to the measured energies and
free-electron band structure. This model reproduces correctly the
density of states, effective masses, and deformation potentials,
without taking into account the interactions between more distinct,
e.g., next NN ions.

The (Ga,Mn)As MBE-grown layers are diluted ferromagnetic semiconductors,
with the Curie temperature which depends on both the Mn magnetic ions
content and the concentration of holes in the valence band.  The
valence-band structure of (Ga,Mn)As with small fraction of Mn was
shown to be quite similar to that of GaAs \cite{okabayashi} and we
take most of parameters to be identical to those in GaAs. The presence
of the Mn ions in the lattice results, however, in spin splittings
of the conduction and valence bands, due to $sp-d$ exchange
interactions between the spins of the band electrons and localized Mn
magnetic moments. These interactions are included into the
tight-binding Hamiltonian using the mean-field prescription with the
experimental values of the exchange integrals $N_0\beta=-1.2$~eV and
$N_0\alpha=0.2$~eV.\cite{okabayashi}

We built the SL assuming that the band offsets at the (Ga,Mn)As and
GaAs interfaces are induced solely by the spin splittings in the
(Ga,Mn)As bands. In structures incorporating (Al,Ga)As nonmagnetic
layers large band offsets (e.g., for $30\%$ of Al, $0.41$~eV in the
valence and $0.15$~eV in the conduction band) have to be taken into
account. The relatively small lattice mismatch between GaAs and
(Ga,Mn)As, \cite{kuryliszyn_jap} as well as the strains resulting from
it, have been ignored.  All the experimentally studied (Ga,Mn)As-based
SL were grown on GaAs substrate along [001] crystallographic axis. In
this case the primitive lattice vectors, which define the SL
elementary cell are: ${\bf{a_1}}=a\sqrt{3}/2[1,1,0]$;
${\bf{a_2}}=a\sqrt{3}/2[1,0,m+n]$; ${\bf{a_3}}=a\sqrt{3}/2[0,1,m+n]$.
The spins in the magnetic layers are aligned along the [100]
direction.\cite{kepar} In order to calculate IEC in the spirit of
Ref.~\onlinecite{blin}, one has to compare the total energy of the
valence electrons for two different SL, one with parallel and the
other with antiparallel spin alignment in consecutive magnetic layers.
Thus, the SL elementary magnetic cell, which has to be considered,
must contain at least two magnetic layers, i.e., it should consists of
$2(n+m)$ monolayers.  This together with the used description of the
constituent materials leads to $80(m+n) \times 80(m+n)$ matrix for the
SL tight-binding Hamiltonian.  After the numerical diagonalization of
the two Hamiltonian matrices, which correspond to the two different
relative spin configurations of the (Ga,Mn)As FM layers, the SL
occupied states' energies were summed up to the Fermi energy and
integrated over the entire Brillouin zone. The position of the Fermi
level in the SL valence band is assumed to be determined by the
average number of holes present in the structure -- for (Ga,Mn)As/GaAs
it is given by $\frac{a^3}{4}(p_m \cdot m)$, whereas for
(Ga,Mn)As/GaAs:Be by $\frac{a^3}{4}(p_m \cdot m + p_n \cdot n)$.  In
(Ga,Mn)As/(Al,Ga)As with high Al content the holes are confined in the
(Ga,Mn)As layers, due to the high potential barriers introduced in the
valence band by the spacer layer.  As all the studied structures
contain (Ga,Mn)As layers which were not annealed, we assume the hole
density in (Ga,Mn)As to be equal to $p_m = 2\times 10^{20}$~cm$^{-3}$
for the sample with $4\%$ of Mn and $p_m= 3\times 10^{20}$~cm$^{-3}$
for the sample $6\%$ of Mn.\cite{osiny} The density of holes
introduced by Be in the spacer is assumed to be $p_n = 1.21\times
10^{20}$~cm$^{-3}$.  \cite{lee}

The strength of the interlayer magnetic coupling is given by the
difference $\Delta E$ between the energies of valence electrons in SL
calculated for the two spin configurations, per unit surface of the
layer.  The preferred spin configuration in consecutive magnetic
layers is given by the sign of $\Delta E$ -- the negative value
corresponds to FM IEC whereas the positive sign indicates a AFM
correlation. The results of the calculations are summarized in the
Figs.~1--3.  Like for the EuS-based structures, here again $J$
practically does not depend on the thickness of the magnetic layer --
all the presented results are calculated for $m=4$.

\begin{figure}
\epsfig{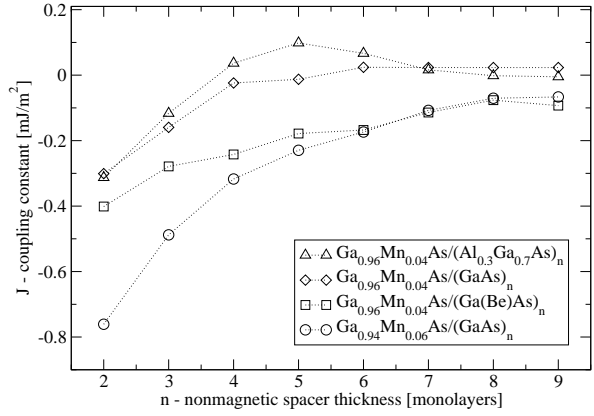}
\caption {The interlayer exchange coupling calculated for (Ga,Mn)As-based structures,
  which were studied experimentally in Refs. \onlinecite{akiba},
  \onlinecite{kepar} and \onlinecite{lee}.}
\label{J}
\end{figure}
In Fig.~\ref{J} the calculated dependence of the interlayer coupling
constant $J=\Delta E/4$ on the spacer thickness $n$ for
Ga$_{0.94}$Mn$_{0.06}$As/GaAs SL is shown together with the results
obtained for Ga$_{0.96}$Mn$_{0.04}$As/GaAs, without and with Be-doping
(the latter introducing $p_n$ holes in the spacer layer) and
Ga$_{0.96}$Mn$_{0.04}$As/Ga$_{0.7}$Al$_{0.3}$As, i.e., for the other
experimentally studied (Ga,Mn)As-based structures.  In qualitative
agreement with the experiment, the obtained IEC for all these
structures is, in principle, FM and decreases  with the thickness of nonmagnetic layers.
The higher the hole concentration in the SL the
stronger is the ICE. For the (Ga,Mn)As/(Al,Ga)As sample, where the holes are
confined in the deep wells formed by the barriers of spacer layers,
the IEC is considerably suppressed and vanishes for $n>7$, as measured
in Ref.~\onlinecite{akiba} for the structure with $n=10$. This result
does not confirm, however, that there is no IEC without holes in the
spacer layer. For very thin spacers, 2-3
monolayers, a strong FM coupling, and for $n=5$ an AFM coupling was
obtained (see Fig.~\ref{J}).

\begin{figure}
\epsfig{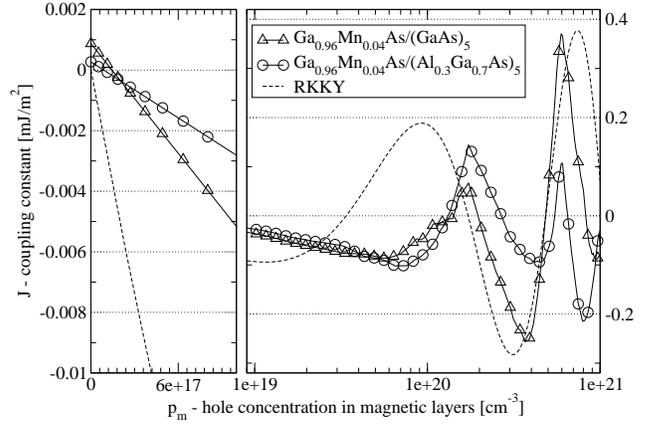}
\caption {The calculated dependence of interlayer coupling constant 
  $J$ on the hole concentration in SL consisting of
  alternating $m=4$ Ga$_{0.96}$Mn$_{0.04}$As monolayers and $n=5$
  monolayers of GaAs or Ga$_{0.7}$Al$_{0.3}$As. $J_{RKKY}$ is shown
  for comparison. }
\label{kon}
\end{figure}
\begin{figure}
\epsfig{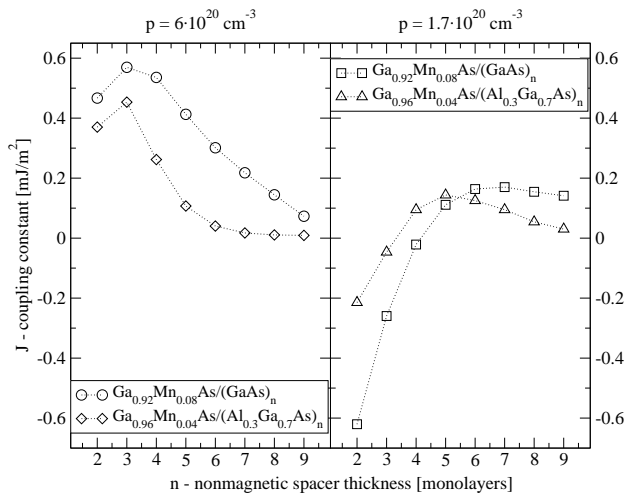}
\caption{The coupling constant vs. spacer thickness for (Ga,Mn)As-based SL in the
two regions of hole concentration, in which the model predicts an antiferromagnetic interlayer
coupling.}
\label{predicts}
\end{figure}

To make these results and the role played by holes more clear, the
dependence of the calculated interlayer coupling constant $J$ on the
position of the Fermi level, i.e., on the average concentration of
holes in the SL valence band, was studied.  As shown in
Fig.~\ref{kon}, $J$ has an oscillatory RKKY-like character (for
comparison IEC mediated by RKKY interaction, i.e., 
$J_{RKKY}\sim {k_F^2} F(2k_F r)$, where $k_F$ is the
Fermi wave vector and $F(x) = (x \cos x - \sin x)/{x^2}$,\cite{yafet}
is presented in the figure by the dashed line). In contrast to
$J_{RKKY}$, at the zero hole concentration limit $J$ tends not to
zero, but to a finite positive value, which corresponds to IEC
mediated by valence band electrons in a hypothetical (Ga,Mn)As/GaAs SL
with completely filled valence bands.  In (Ga,Mn)As/(Al,Ga)As SL, for
the concentrations up to about $4 \times 10^{20}$~cm$^{-3}$ the holes
are confined in the wells -- when the Fermi level reaches the value of
the band offset between (Ga,Mn)As and (Al,Ga)As, the distribution of
holes in the SL changes and the obtained $J$ values for higher
concentrations do not follow the previous trends.  Importantly, as
suggested before in Ref.~\onlinecite{jungwirth}, the presented in
Fig.~\ref{kon} results indicate that in (Ga,Mn)As-based
heterostructures also the AFM coupling between FM layers could be
achieved by an appropriate engineering of the SL and a proper choice
of constituent materials. On the grounds of the presented results, structures
particularly suitable for the observation of AFM correlations can be
suggested.  These seem to be SL in which the hole concentration is
either increased (e.g., by appropriate annealing during the MBE growth
of the SL) to about $6 \times 10^{20}$~cm$^{-3}$ or kept as low as
$1.5-2.5 \times 10^{20}$~cm$^{-3}$. It should be noted that in the
former one can expect also high Curie temperature. The
(Ga,Mn)As/(Al,Ga)As system is additionally interesting because here,
due to high potential barriers in the nonmagnetic spacers, the
carriers are confined in the DMS layers, what can result in strongly
spin-polarized charge density.  In the latter heterostructures the hight of
the barrier, i.e., the Al content, is very important -- the results
for (Ga,Mn)As/AlAs SL (for clarity not included in the figure) show
that very high barriers reduce extremely the IEC in both, FM and AFM,
regions. Finally, in Fig.~\ref{predicts} we show the dependencies of
$J$ on the thickness of the spacer layer $n$ for the
Ga$_{0.92}$Mn$_{0.08}$As/GaAs and
Ga$_{0.96}$Mn$_{0.04}$As/Al$_{0.3}$Ga$_{0.7}$As SL with appropriate
for AFM IEC hole concentrations. For the higher concentration the
coupling is stronger for both structures but it decreases more rapidly
with the spacer thickness. It should be noted that SL with the spacers as thin as 3
monolayers, for which the strongest coupling has been predicted, would
be difficult to obtain, due to the strong
interdiffusion in the LT MBE grown (Ga,Mn)As
structures. \cite{sadowski} Still, for $n=5-6$, the predicted AFM IEC is
of the same order of magnitude as the FM coupling observed in the
(Ga,Mn)As-based SL.

In conclusion,  we have studied, within a tight binding model, the sensitivity of the band structure
of (Ga,Mn)As-based SL to the spin configuration in successive DMS
layers. Such effects describe correctly the AFM IEC between the FM
layers in EuS/PbS and EuS/YbSe and are, up to now, the only
effective mechanism capable to explain the origin of interlayer
correlations in AFM EuTe/PbTe SL. We have shown that by this
mechanism also the FM interlayer coupling in (Ga,Mn)As/GaAs SL can be
described. Moreover, the model points to a
possibility of engineering (Ga,Mn)As-based multilayers for obtaining
an AFM interlayer coupling.

{\bf Acknowledgments:} The authors thank T. Story for
elucidating discussions. This work was supported by the
Polish Ministry of Science (PBZ-KBN-044/P03/2001) and FENIKS
(EC:G5RD-CT-2001-00535) projects.

\end{document}